# A Taxonomy of Workflow Management Systems for Grid Computing

Jia Yu and Rajkumar Buyya

**Gri**d Computing and **D**istributed **S**ystems (GRIDS) Laboratory
Department of Computer Science and Software Engineering
The University of Melbourne, Australia
{jiayu, raj}@cs.mu.oz.au

## ABSTRACT

With the advent of Grid and application technologies, scientists and engineers are building more and more complex applications to manage and process large data sets, and execute scientific experiments on distributed resources. Such application scenarios require means for composing and executing complex workflows. Therefore, many efforts have been made towards the development of workflow management systems for Grid computing. In this paper, we propose a taxonomy that characterizes and classifies various approaches for building and executing workflows on Grids. We also survey several representative Grid workflow systems developed by various projects world-wide to demonstrate the comprehensiveness of the taxonomy. The taxonomy not only highlights the design and engineering similarities and differences of state-of-the-art in Grid workflow systems, but also identifies the areas that need further research.

## 1. INTRODUCTION

Grids [50] are emerging as a global cyber-infrastructure for the next-generation of e-Science applications by integrating large-scale, distributed and heterogeneous resources. Scientific communities, ranging from high-energy physics, gravitational-wave physics, geophysics, astronomy to bioinformatics, are utilizing Grids to share, manage and process large data sets. In order to support complex scientific experiments, distributed resources such as computational devices, data, applications, and scientific instruments need to be orchestrated along with managing the application workflow operations within Grid environments [87].

Grid workflow can be seen as a collection of tasks that are processed on distributed resources in a well-defined order to accomplish a specific goal. Workflow management techniques have been developed for over 20 years, especially in business management and office automation, and production management [5]. Many successful approaches can be applied to Grid workflow for scientific applications. However, there exist several differences between Grid-based scientific workflows and conventional workflows [84][25] [107] such as long lasting workflow execution, large data flow, heterogeneous resources, multiple administrative domains, and dynamic resource availability and utilization.

Workflow management systems that take care of defining, managing and executing Grid workflows are increasingly being utilized for a large range of scientific applications. The workflow paradigm for applications composition on Grids offers several advantages, such as [112] :
- Ability to build dynamic applications which orchestrate distributed resources.
- Promotion of inter-organization collaborations.
- Utilizing resources that are located in a particular domain to increase throughput or reduce execution costs.
- Execution spanning multiple administrative domains to obtain specific processing capabilities.
- Integration of multiple teams involved in management of different parts of the experiment workflow.

Figure 1 shows architecture and functionalities supported by various components of the Grid workflow system which is based on the workflow reference model [34] proposed by Workflow Management Coalition (WfMC) [132] in 1995. At the highest level, functions of Grid workflow management systems could be characterized into *build time* functions and *run time* functions. The build-time functions are concerned with defining, and modeling workflow tasks and their dependencies; while the run-time functions are concerned with managing the workflow execution and interactions with Grid resources for processing workflow applications. Users interact with workflow modeling tools to generate a workflow specification, which is submitted to a run-time service called workflow enactment service for execution. The major functions provided by workflow enactment services are scheduling, fault management and data movement. A workflow enactment service may be built on the top of low level Grid middleware (e.g. Globus toolkit [58], UNICORE [123] and Alchemi [81]), through which the workflow management systems invoke services provided by Grid resources. At both build-time and run-time stages, the information about resources and applications may need to be retrieved.

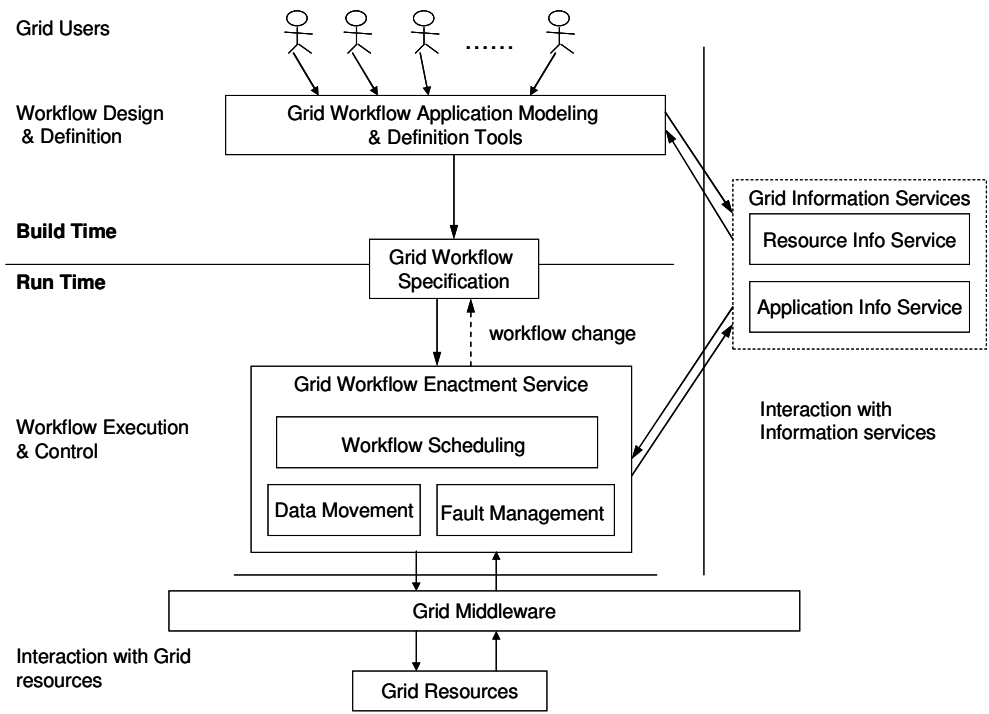

*Figure 1*. Grid Workflow Management System.

In the recent past, several Grid workflow systems have been proposed and developed. In order to enhance understanding of the field, we propose a taxonomy that primarily (a) captures architectural styles and (b) identifies design and engineering similarities and differences between them. The taxonomy provides an in-depth understanding of building and executing workflows on Grids. There are a number of proposed taxonomies for distributed and heterogeneous computing such as [29][103][20][72]. However, none of these focuses on workflow management. The proposed taxonomy, presented in Section 2, classifies approaches based on major functions and architectural styles of Grid workflow systems. We also survey, in Section 3, several Grid workflow management systems along with mapping the taxonomy and identifying the areas that need further investigation.

## 2. TAXONOMY

The taxonomy characterizes and classifies approaches of workflow management in the context of Grid computing. As shown in Figure 2, the taxonomy consists of sub-taxonomies based on the five elements of a Grid workflow management system: (a) workflow design, (b) information retrieval, (c) workflow scheduling, (d) fault tolerance and (e) data movement. In this section, we look at each element and its taxonomy in detail.

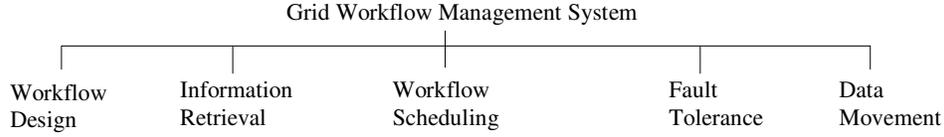

*Figure 2*. Elements of a Grid Workflow Management System.

### 2.1 Workflow Design

The workflow design includes key factors involved in the workflow at build-time. It consists of four sub-taxonomies, namely (a) workflow structure, (b) workflow model/specification, (c) workflow composition system, and (d) workflow QoS (Quality of Service) constraints.

*2.1.1 Workflow Structure*

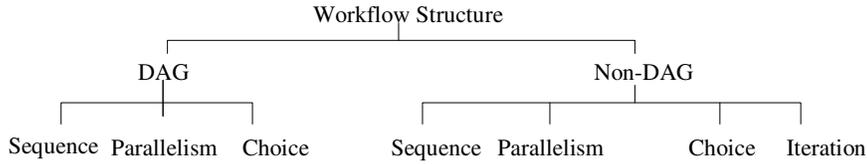

*Figure 3*. Workflow Structure Taxonomy.

A workflow is composed by connecting multiple tasks according to their dependencies. Workflow structure, also referred as workflow pattern [2][3][6], indicates the temporal relationship between tasks. Figure 3 shows workflow structure taxonomy. In general, the workflow can be represented as a Directed Acyclic Graph (DAG) and non-DAG.

In DAG-based workflow, workflow structure can be categorized into sequence, parallelism, and choice. *Sequence* is defined as an ordered series of tasks, with one task starting after a previous task has completed. *Parallelism* represents tasks which are performed concurrently, rather than serially. In *choice* control pattern, a task is selected to execute at run-time when its associated conditions are true.

In addition to all patterns contained in a DAG-based workflow, non-DAG workflow also includes *iteration* structure, in which sections of workflow tasks in an iteration block are allowed to be repeated. Iteration is also known as *loop* or *cycle*. Iteration structure is quite frequent in scientific applications, where one or more tasks needed to be executed repeatedly [86].

These four types of workflow structure, namely *sequence*, *parallelism*, *choice* and *iteration*, can be used to construct many complex workflows. Moreover, sub-workflows can also use these types of workflow structure as building blocks to form a large-scale workflow.

*2.1.2 Workflow Model/Specification*

Workflow Model (also called workflow specification) defines a workflow including its task definition and structure definition. As shown in Figure 4, there are two types of workflow models, namely *abstract* model and *concrete* model. They are also denoted as abstract workflow and concrete workflow [39][41]. In some literature such as [80], *concrete* model is also referred as executable workflow.

In the *abstract* model, a workflow is described in an abstract form, in which the workflow is specified without referring to specific Grid resources for task execution. The *abstract* model provides a flexible way for users to define workflows without being concerned about low-level implementation details. Tasks in the *abstract* model are portable and can be mapped onto any suitable Grid services at run-time by using suitable discovery and mapping mechanisms. The *abstract* model also eases the sharing of workflow descriptions between Grid users [41]; in particular it benefits the participants of Virtual Organizations (VOs) [51].

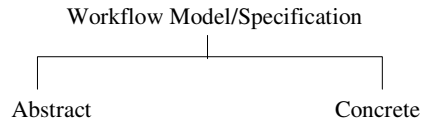

*Figure 4*. Workflow Model Taxonomy.

In contrast, the *concrete* model binds workflow tasks to specific resources. In some cases, the *concrete* model may include nodes acting as data movement to stage data in and out of the computation and data publication to publish newly derived data into VO [41]. In another situation, tasks in the *concrete* model may also include necessary application movement to transfer computational code to the data site for large scale data analysis.

Given the dynamic nature of the Grid environment, it is more suitable for users to define workflow applications in the *abstract* model. A full or partial *concrete* model can be generated just before or during workflow execution according to the current status of resources. Additionally, in some systems [139], every task in the workflow is concretized only at the time of task execution. However, *concrete* models may be needed by some end users who want to control the execution sequence [74].

*2.1.3 Workflow Composition System*

Workflow composition systems are designed for enabling users to assemble components into workflows. They need to provide a high level view for the construction of Grid workflow applications and hide the complexity of underlying Grid systems. Figure 5 shows the taxonomy for the workflow composition systems. *User-directed* composition systems allow users to edit workflows directly, whereas *automatic* composition systems generate workflows for users automatically. In general, users can use workflow languages for *language-based modeling* and the tools for *graph-based modeling* to compose workflows.

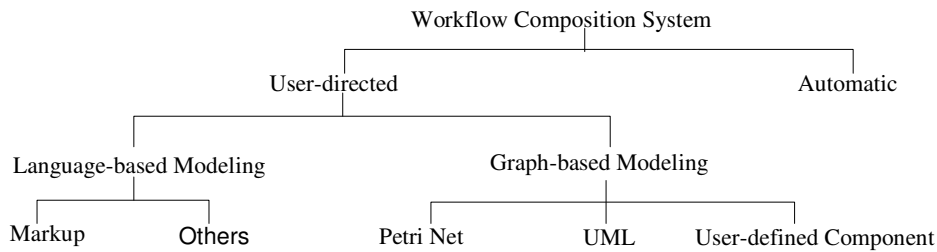

*Figure 5*. Workflow Composition System Taxonomy.

Within *language-based modeling*, users may express workflow using a *markup* language such as Extensible Markup Language (XML) [127] (e.g. GridAnt [74], WSFL [76], XLANG [120], BPEL4WS [14], W3C XML-Pipeline language, and Gridbus Workflow [139]) or other formats (e.g. Condor DAGman [115]). *Language-based modeling* may be convenient for skilled users, but they require users to memorize a lot of language-specific syntax. In addition, it is impossible for users to express a complex and large workflow by scripting workflow components manually. However, workflow languages are more appropriate for storing and transfer as compared to graphical representation which required to be converted into other form for such manipulation. So in most Grid systems, workflow languages are designed to bridge the gap between the graphical clients and the Grid workflow execution engine [61]. XML-based languages are preferred to other scripting languages as XML provides a set of rules for users to describe information in nested structures. Moreover, the syntax of XML-based languages provided through DTD (Document Type Definition) [127] and XML schema [130] can be used to validate user inputs. Furthermore, many XML parsing tools (e.g. JDOM [68] and dom4j [43] ) are widely available.

*Graph-based modeling* allows graphical definition of an arbitrary workflow through a few basic graph elements. It allows users to work with a graphical representation of the workflow. Users can compose and review a workflow by just clicking and dropping the components of interest. It avoids low-level details and hence enables users to focus on higher levels of abstraction at application level [63]. The major modeling approaches are *Petri Nets* [99], *UML* (Unified Modeling Language) [94] and *user-defined component*. *Graph-based modeling* is more preferred by users as opposed to *language-based modeling*.

*Petri Nets* are a special class of directed graphs that can model sequential, parallel, loops and conditional execution of tasks [61][64]. They have been used in many workflow management systems such as Grid-Flow [61], FlowManager [75], and XRL/Flower [126]. *UML* activity diagrams [97] have also been extended and applied to be a workflow specification language [17][44][100]. Compared with *UML* activity diagrams, *Petri Nets* have formal semantics and have been used widely for constructing several workflows [1][45]. A vast number of algorithms and tools for *Petri Nets* analysis have been developed along the years [84]. However, Eshuis et. al [45] argues that *Petri Nets* may be unable to model workflow activities accurately without extending its semantics and this drawback has been addressed in *UML* activity diagrams. Rather than following the standard syntax and semantics of *Petri Nets* and *UML*, many workflow editors for Grid workflow tools create their own graphical representation of workflow components. For example, Triana [118] allows users to predefine software components and reuse them to design DAG-based workflows. Kepler [12] provides graphical environment and a framework that supports the design and reuse of grid workflows. These tools are more convenient for users to manipulate their workflow applications, as they provide a more user-friendly programming environment. They have also been integrated into underlying local applications, Grid middleware and monitoring systems. For example, P-GRADE [79][70] interoperates with a wide range of parallel applications in addition to Condor and Globus based Grid middleware. It also allows users to access and modify program code of a workflow task through the graphical editor. However, lack of standards hinders the collaboration between these projects. Many works are thus replicated such as different user interfaces developed by different projects for the same functionality. Moreover, workflow structures supported by most of them are limited to only sequence and parallelism.

Graphical representation is very intuitive and can be handled easily even by a non-expert user. However, the layout of workflow components on a display screen can become very huge and difficult to manage [96]. One of solutions to overcome this limitation is to use hierarchical graph definition [64]. Another solution is to have a system which composes workflows automatically. Pegasus [41] is one of automatic composition systems for Grid computing and it has to be adapted to particular applications, because the composition is based on application-dependent metadata. It receives a metadata description of desired data products and initial input values from users. The tasks are then composed automatically to form a workflow by querying a virtual data catalog [52] that contains information for data derivation of application components. Compared with *user-directed* systems, *automatic* composition systems are ideal for large scale workflows which are very time consuming to compose manually. However, *automatic* composition of application components is challenging because it is difficult to capture the functionality of components and data types used by the components [96][27].

*2.1.4 Workflow QoS Constraints*

In Grid environments, there are a large number of similar or equivalent resources provided by different third parties. Grid users can select suitable resources and use them for their workflow applications. These resources may provide the same functionality, but optimize different QoS measures. In addition, different users or applications may have different expectations and requirements. Therefore, it is not sufficient for a workflow management system to only consider functional characteristics of the workflow. QoS requirements such as time limit (deadline) and expenditure limit (budget) for workflow execution also need to be managed by workflow management systems. Users must be able to specify their QoS expectations of the workflow at the design level. Then, actions conducted by workflow systems using run-time must be chosen according to the initial QoS requirements.

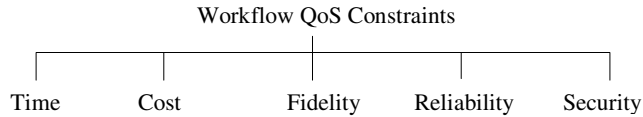

*Figure 6*. Workflow QoS Constraints Taxonomy.

Figure 6 shows the taxonomy of Grid workflow QoS constraints based on a QoS model for Web services based workflow provided by Cardoso et al [28] and QoS of Web services [83][98]. It includes five dimensions: *time*, *cost*, *fidelity*, *reliability* and *security*. *Time* is a basic measure of performance. For workflow systems, it refers to the total time required for completing the execution of a workflow. *Cost* represents the cost associated with the execution of workflows including the cost for managing workflow systems and usage charge of Grid resources for processing workflow tasks. *Fidelity* refers to the measurement related to the quality of the output of workflow execution. *Reliability* is related to the number of failures for the execution of workflow task execution. *Security* refers to confidentiality of the execution of workflow tasks and trustworthiness of resources.

As indicated in Figure 7, there are two different ways to assign QoS constraints in a workflow model. One way is to allow users to assign QoS constraints at *task-level*. The overall QoS can be assessed by computing all individual tasks. For example, a user assigns desired execution time for every task in a workflow. The deadline for the entire workflow execution can be calculated by a workflow reduction algorithm (e.g. SWR(w) algorithm [26]). Another way is to assign QoS constrains at *workflow-level*, allowing users to define the overall workflow QoS requirements. However, QoS constraints for each task may be required by schedulers for resource allocation at run-time. In the example of time dimension, users may like to specify a deadline for the entire workflow execution rather than for every single task. In order to fulfill the deadline for the entire workflow, the scheduler needs to decide how fast each task has to be processed using a deadline assignment approach (e.g. UD, ED, EQS and EQF strategies in [71]).

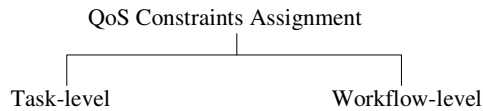

*Figure 7*. QoS Constraints Assignment Taxonomy.

**2.2 Information Retrieval**

A Grid workflow management system does not execute the tasks itself, but it merely coordinates the execution of the tasks by the Grid resources. To map tasks onto suitable resources, information about the resources has to be retrieved from appropriate entities [136]. As indicated in Figure 8, there are three dimensions of information retrieval: *static information*, *historical information* and *dynamic information*.

*Static information* refers to information that does not vary with time. It may include *infrastructure-related* (e.g. the number of processors), *configuration-related* (e.g. operating system, libraries), *QoS-related* (e.g. flat usage charge), *access-related* (e.g. service operations), and *user-related* information (e.g.

authentication ID). Generally, *static information* is utilized by Grid workflow management systems to pre-select resources during the initiation of the workflow execution.

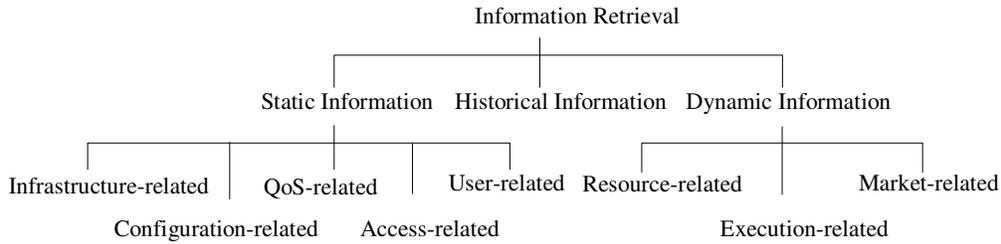

*Figure 8*. Information Retrieval Taxonomy.

As Grid resources are not dedicated to the owners of the workflow management systems, the Grid workflow management system also needs to identify *dynamic information* such as resource accessibility, system workload, and network performance during execution time. Unlike *static information*, *dynamic information* reflects the status of the Grid resources, such as load average of a cluster, available disk space, CPU usage, and active processes. It also includes task execution information and market related information such as dynamic resource price.

*Historical information* is obtained from previous events that have occurred such as performance history and execution history of Grid resources and application components. Generally, workflow management systems can analyze historical information to predict the future behaviors of resources and application components on a given set of resources. Historical information can also be used to improve the reliability of future workflow execution. For example, the user can correct the logic of a failed workflow according to the log of the workflow system.

Several information services are available for accessing static and dynamic information about Grid resources. For example, Monitoring and Discovery System (MDS) [104] provides static hardware information such as CPU type, memory size and software information such as operating system information, and some dynamic information such as CPU load snapshot. Network Weather Service (NWS) [131] provides additional dynamic information about availability of CPU, memory, and bandwidth. An object oriented model for publication and retrieval of electronic resources is given in [32].

**2.3 Workflow Scheduling**

Casavant et al. [29] categorized task scheduling in distributed computing systems into 'local' task scheduling and 'global' task scheduling. Local scheduling involves handling the assignment of tasks to time-slices of a single resource whereas global scheduling involves deciding where to execute a task. According to this definition, workflow scheduling is a kind of global task scheduling as it focuses on mapping and managing the execution of inter-dependent tasks on shared resources that are not directly under its control.

The workflow scheduler needs to coordinate with diverse local management systems as Grid resources are heterogeneous in terms of local configurations and local policies. Taking into account users' QoS constraints is also important in the scheduling process so as to satisfy user requirements. In this section, we discuss workflow scheduling taxonomy from the view of (a) scheduling architecture, (b) decision making, (c) planning scheme, (d) scheduling strategy, and (e) performance estimation.

*2.3.1 Scheduling Architecture*

The architecture for the scheduling infrastructure is very important for the scalability, autonomy, quality and performance of the system [62]. Three major categories of workflow scheduling architecture as shown in Figure 9 are centralized, hierarchical and decentralized scheduling schemes.

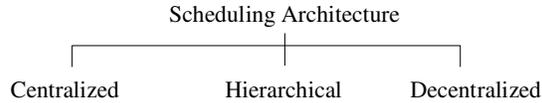

*Figure 9.* Scheduling Architecture Taxonomy.

In the *centralized* workflow enactment environment, one central workflow scheduler makes scheduling decisions for all tasks in the workflow. The scheduler has information of the entire workflow and collects information of all available processing resources. It is believed that the centralized scheme can produce efficient schedules because it has all necessary information [62]. However, it is not scalable with respect to the number of tasks, the classes and number of Grid resources that are generally autonomous. It is thus only suitable for a small scale workflow or a large scale workflow in which every task has the same objective (e.g. same class of resources).

For *hierarchical* scheduling, there is a central manager and multiple lower-level sub-workflow schedulers. The manager is responsible for controlling the workflow execution and assigning the sub-workflows to the low-level schedulers. For example, in GridFlow project [25], there is one workflow manager and multiple lower-level schedulers. The workflow manager schedules sub-workflows onto corresponding lower-level schedulers. Each lower-level scheduler is responsible for scheduling tasks in a sub-workflow onto resources owned by one organization. The major advantage of using the hierarchical architecture is that the different scheduling policies can be deployed in the central manager and lower-level schedulers [62]. However, the failure of the central manager will result in entire system failure. In contrast, there is no central controller in *decentralized* scheduling. Every scheduler can communicate each other and schedule a sub-workflow to another scheduler with lower load.

Unlike *centralized* scheduling, both *hierarchical* and *decentralized* scheduling allow tasks to be scheduled by multiple schedulers. Therefore, one scheduler only maintains the information related to a sub-workflow. Thus, compared to *centralized* scheduling, they are more scalable since they limit the number of tasks managed by one scheduler. However, the best decision made for a partial workflow may lead to a worse performance for the overall workflow execution. Moreover, conflict problems are more severe [85]. One example of conflict is tasks from different sub-workflows scheduled by different schedulers may compete for the same resource. Compared to the hierarchical scheme, the decentralized scheme is more scalable but faces more challenges to generate optimal solutions for overall workflow performance and minimize conflict problems.

*2.3.2 Decision Making*

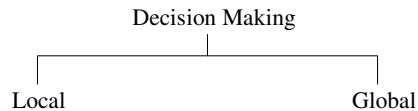

*Figure 10.* Decision Making Taxonomy.

It is difficult to find a single best solution of mapping workflow onto resources for all workflow applications, since the applications can have very different characteristics. It depends to some degree on the application models scheduled. In general, decisions about mapping tasks in a workflow onto resources can be based on the information of current task or entire workflow, namely *local* decision and *global* decision [39] as shown in Figure 10. Scheduling decision made with reference to just the task or sub-workflow at hand is called *local* decision while scheduling decision made with reference to the whole workflow is called *global* decision.

*Local* decision only takes one task or sub-workflow into account, so it may produce a best schedule for current task or sub-workflow but could reduce the entire workflow performance. An example given by Deelman et al [39] assumes there is a data-intensive application where the overall run-time is driven by data transfer costs. Consider the case where the output of a task is significantly bigger than its input and the resources capable of processing its children tasks are in limited locations. If the resource selection is only based on local decision, the selected resource is most suitable for the task, but can be further than other alternatives closer to the children processing nodes. Therefore, it will lead to significant higher transfer cost.

Scheduling workflow tasks using *global* decision improves the performance of entire workflow. There are some algorithms of scheduling task graphs in parallel systems that could be applied to Grid workflow scheduling. Li et. al [77] developed Forward-Looking Analysis Method (FLAM). It analyses dependencies of the entire graph to resolve the conflicts of parallel tasks which are competing for the same resource. It is believed that the *global* decision based scheduling can provide better overall result. However, it may take much more time in scheduling. Thus, the overhead produced by global scheduling will reduce the overall benefit and can even exceed the benefits it will produce [39]. Therefore, the decision of scheduling should not be made without considering the balance between the overall execution time and scheduling time. However, for some applications such as a data analysis application where the outputs of tasks in the workflow are always smaller than the inputs, using *local* decision based scheduling is sufficient.

*2.3.3 Planning Scheme*

It is concerned with a scheme for translating abstract workflows to concrete workflows. As shown in Figure 11, schemes for the schedule planning of workflow applications can be categorized into either *static* scheme or *dynamic* scheme. In *static* scheme, concrete models have to be generated before the execution according to current information about execution environment and dynamically changing state of the resources is not taken into account. In contrast, *dynamic* scheme uses both dynamic information and static information about resources to make scheduling decisions at run-time.

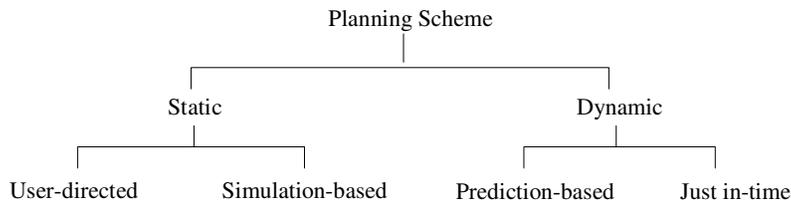

*Figure 11*. Planning Scheme Taxonomy.

*Static* scheme, also known as *full-ahead* planning, includes *user-directed* and *simulation-based* scheduling. In *user-directed* scheduling approach, users emulate the scheduling process and make resources mapping decision according to their knowledge, preference and/or performance criteria. For example, users prefer to map tasks to resources on which he has not experienced failures. In the *simulation-based* scheduling approach, a 'best' schedule is achieved by simulating task execution on a given set of resources before a workflow starts execution. The simulation can be processed based on static information or the result of performance estimation. For example, in GridFlow [25], the 'best' resource selected for scheduling a task is based on the predictive task execution time that resource provides.

*Dynamic* scheme includes *prediction-based* and *just in-time* scheduling. *Prediction-based* dynamic scheduling uses dynamic information, which is available only at execution time, in conjunction with some results based on prediction. It is similar to *simulation-based* static scheduling, in which they predict the performance of task execution on resources and generate a near optimal schedule for the task before it starts execution. However, it changes the initial schedule dynamically during the execution. For example, GrADS [31] generates preliminary mapping by using prediction results, but it migrates a task execution to another resource when its initial contract is broken or a better resource is found for execution. Sakellariou et al [105] developed a low-cost rescheduling policy for the mapping of workflows on Grids. It considers rescheduling workflow tasks at a few carefully selected points during execution in a dynamically changing Grid environment since the initial schedule built using inaccurate predictions that affects performance significantly.

Rather than making a schedule ahead, *just in-time* scheduling [41] only makes scheduling decision at the time of task execution. Planning ahead in Grid environments may produce a poor schedule, since Grid is a dynamic environment where utilization and availability of resources varies over time and a better resource can join at any time. Moreover, it is not easy to predict execution time accurately of all application components on Grid resources. However, as the technologies of advance reservations [114] for various resources improve, it is believed that the role of static and prediction-based planning will increase [39].

*2.3.4 Scheduling Strategy*

In general, scheduling workflow applications in a distributed system is an NP-complete problem [49]. Therefore, many heuristics have been developed to obtain near-optimal solutions to match users' QoS constraints. As shown in Figure 12 we categorize strategies of major scheduling approaches into *performance-driven*, *market-driven* and *trust-driven*.

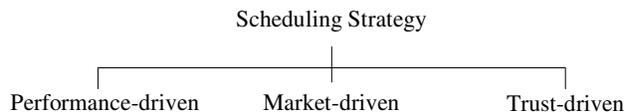

*Figure 12*. Scheduling Strategy Taxonomy.

The *performance-driven* strategy tries to find a mapping of workflow tasks onto resources that achieves optimal execution performance such as minimize overall execution time. Most of Grid workflow scheduling systems falls in this category. GrADS [31] optimizes DAG-based workflows using Min-Min, Max-Min and Suffrage heuristics, hoping to obtain minimum completion times. Prodan et.al [101] uses classical genetic algorithms with cycle elimination techniques to minimize non-DAG based workflow execution on Grids.

The *market-driven* strategy employs market models to manage resource allocation for processing workflow tasks. It applies computational economy principle and establishes an open electronic marketplace between workflow management systems and participating resource providers. Workflow schedulers act as consumers buying services from the resource providers and pay some notion of electronic currency for executing tasks in the workflow. The tasks in the workflow are dynamically scheduled at run-time depending on resource cost, quality and availability, to achieve the desired level of quality for deadline and budget. Unlike the *performance-driven* strategy, *market-driven* schedulers may choose a resource with later deadline if its usage price is cheaper. The *market-driven* strategy has been applied into several Grid systems such as Nimrod-G [21] and Gridbus data resource broker [125]. One example of the *market-driven* workflow scheduling proposed by Geppert et al [57] utilizes market mechanisms during the task assignment. In the system, bids are collected from eligible resource providers for each task. The optimal bid is selected by computing the amount of time and cost saved or overdrawn up to the point. If the execution time has been minimized at the expense of an overdrawn cost, a bid with lower price will be chosen as the optimal bid. Consequently, scheduler assigns the task to the resource whose provider offers the optimal bid. A recent work on cost-based scheduling of workflow tasks on Grids is reported in [19].

Recently *trust-driven* scheduling approaches (e.g. CCOF project in [141] and GridSec project in [110][109]) in distributed systems are emerging. *Trust-driven* schedulers select resources based on their trust levels. For example, within GridSec, the scheduler accesses the trust level of Grid sites. It maps tasks onto resources whose trust level is higher than users' demand. Trust model of resources is based on attributes such as security policy, accumulated reputation, self-defense capability, attack history, and site vulnerability. By using *trust-driven* approaches, workflow management systems can reduce the chance of selecting malicious hosts, and non-reputable resources [141]. Therefore, overall accuracy and reliability of workflow execution will be increased.

*2.3.5 Performance Estimation*

In order to produce a good schedule, estimating the performance of tasks on resources is crucial, especially for constructing a preliminary workflow schedule. By using performance estimation techniques, it is possible for workflow schedulers to predict how tasks in a workflow or sub-workflow will behave on distributed heterogeneous resources and thus make decisions on how and where to run their tasks. As indicated in Figure 13, there are several performance estimation approaches: simulation, analytical modeling, historical data, on-line learning, and hybrid approach.

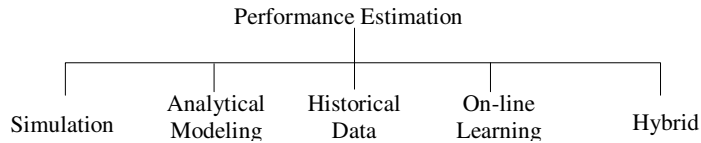

*Figure 13*. Performance Estimation Taxonomy.

The *simulation* approach [42][142] provides resource simulation environments to emulate the execution of tasks in the workflow prior to its actual execution. In the *analytical modeling* approach [31][36][93], a scheduler predicts the performance of tasks in workflow on a given set of resources based on an analytic metric. For example, within GrADS [31], two types of performance models are developed, namely memory hierarchy performance model and computational model. By using these models, one can predict memory requirement and execution time of an application component for a resource according to the associated problem size. The *historical data* approach [86][67][108] relies on historical data to predict the task's execution performance. The historical data related to a particular user's application performance or experience can also be used in predicting the share of available of resources for that user while making scheduling decisions based on QoS constraints. The *on-line learning* approach predicts task execution performance from on-line experience without prior knowledge of the environment's dynamics. For example, Buyya et al. [22] and Galstyan et al. [56] maps a job onto a 'best' Grid resource by learning the completion time of most recent jobs submitted to resources. As historical and on-line learning approaches use experimental data, they can be broadly terms as *empirical modeling* approaches for performance estimation.

In certain conditions, these approaches could be utilized together as a *hybrid* approach for generating performance evaluation of workflow tasks. For instance, Bacigalupo et. al [16] uses both layered queuing modeling and historical performance data to predict the performance of dynamic e-Commerce systems on heterogeneous server. In addition, GrADS constructs computational models semi-automatically by emulating the execution of workflow components on small data sets. That is, it uses a combination of historical and analytical approach for performance estimation.

**2.4 Fault Tolerance**

In Grid environments, resources span across multiple administrative domains and are not under the control of the workflow management systems. Moreover, many users are competing for limited resources. Workflow execution failures may be caused by many reasons, such as the change of resource local policy and the failure of resources and network fabric. Thus, Grid workflow management systems should be able to handle failures flexibly and support reliable executions in the presence of concurrency and failures.

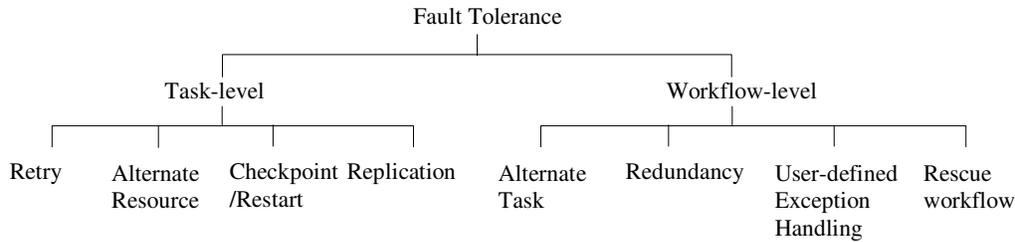

*Figure 14*. Fault Tolerance Taxonomy.

As shown in Figure 14, Hwang et al. [65] divided workflow failure handling techniques into two different levels, namely *task-level* and *workflow-level*. *Task-level* techniques mask the effects of the execution failure of tasks in the workflow, while *workflow-level* techniques manipulate the workflow structure such as execution flow to deal with erroneous conditions.

*Task-level* techniques have been greatly studied in parallel and distributed systems. They can be cataloged into *retry*, *alternate resource*, *checkpoint/restart* and *replication*. The *retry* technique [116] is the simplest failure recovery technique, as it simply tries to execute the same task on the same resource again after failure. The *alternate resource* technique [116] submits failed task to another resource. The *checkpoint/restart* technique [35] moves failed tasks transparently to other resources, so that the task can continue its execution from the point of failure. And the *replication* technique [7][65] runs the same task simultaneously on different Grid resources to ensure task execution provided that at least one of the replicas does not fail.

*Workflow-level* techniques include *alternate task*, *redundancy*, *user-defined exception handling* and *rescue workflow*. The first three approaches proposed in [65] assume there is more than one implementation for a certain computation with different execution characteristics. The *alternate task* technique executes another implementation of a certain task if the previous one failed, while the *redundancy* technique executes multiple alternative tasks simultaneously. The *user-defined exception handling* allows the users to specify a special treatment for a certain failure of a task in workflow. The rescue workflow technique developed in Condor DAGMan system [35] ignores the failed tasks and continues to execute the remainder of the workflow until no more forward progress can be made. Then, a rescue workflow description called rescue DAG, which indicates failed nodes with statistical information, is generated for later submission.

**2.5 Intermediate Data Movement**

For Grid workflow applications, the input files of tasks need to be staged to a remote site before processing the task. Similarly, output files may be required by their children tasks which are processed on other resources. Therefore, the intermediate data has to be staged out to the corresponding Grid sites. Some systems require users to manage intermediate data transfer in the workflow specification, rather than providing *automatic* mechanisms to transfer intermediate data. As indicated in Figure 15, we categorize approaches of *automatic* intermediate data movement into *centralized*, *mediated* and *peer-to-peer*.

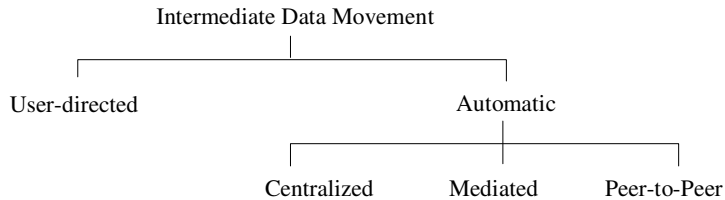

*Figure 15*. Intermediate Data Movement.

Basically the *centralized* approach transfers intermediate data between resources via a central point. For example, the central workflow execution engine can collect the execution results after task completion and transfer them to the processing entities of corresponding successors. The *centralized* approach is easy to implement and suits workflow applications in which large-scale data flow is not required.

Rather than using a central point for the *mediated* approach, the locations of the intermediate data are managed by a distributed data management system. For example, in Pegasus system, the intermediate data generated at every step is registered in a replication catalog service [30], so that input files of every task can be obtained by querying the replication catalog service. The *mediated* approach is more scalable and suitable for applications which need to keep intermediate data for later use.

The *peer-to-peer* approach transfers data between processing resources. Since data is transmitted from the source resource to the destination resource directly without involving any third-party service, it significantly saves the transmission time and reduces the bottleneck problem caused by the *centralized* and *mediated* approaches. Thus, it is suitable for large-scale intermediate data transfer. However, there are more difficulties in deployment because it requires a Grid node to be capable of providing both data management and movement service. In contrast, *centralized* and *meditated* approaches are more suitably used in applications such as bio-applications, in which users need to monitor and browse intermediate results. In addition, they also need to record them for future verification purposes.

## 2. GRID WORKFLOW MANAGEMENT SYSTEM SURVEY

In this section, we present a detailed survey of existing Grid workflow systems in addition to mapping the proposed taxonomy. Table 1 shows the summary of selected Grid workflow management projects. A comparison of various Grid workflow systems and their categorization based on the taxonomy is shown in Table 2, Table 3, and Table 4.

*Table 1*. Summary of Grid Workflow Management Projects.

| Name | Organization | Prerequisite | Grid Integration | Applications | Availability |
|---|---|---|---|---|---|
| DAGMan | University of Wisconsin-Madison, USA. http://www.cs.wisc.edu/condor/dagman/ | Condor | Condor which can run on top of Globus Toolkit version 2 (GT2) | Compute-intensive | GPL(General Public License) |
| Pegasus | University of Southern California, USA. http://pegasus.isi.edu | Condor DAGMan, Globus RLS. | Condor and Globus. | Targeted for data-intensive, but supports other types. | GTPL (Globus Toolkit Public License) |
| Triana | Cardiff University, UK. http://www.trianacode.org/ | Grid Application Toolkit (GAT) | GAT (JXTA, Web services, Globus) | Compute-intensive | BSD License |
| ICENI | London e-Science Centre, UK. http://www.lesc.ic.ac.uk/iceni/ | Globus toolkit | Jini, JXTA, Globus | Compute-intensive | ICENI Open Source Code Licence |
| Taverna | Collaboration between several | Java 1.4+ | Web services, Soaplab, local | Service Grids | GNU Lesser General Public |

| Name | Organization | Prerequisite | Grid Integration | Applications | Availability |
|------|--------------|--------------|------------------|--------------|--------------|
| | European Institutes and industries. http://taverna.sourceforge.net/ | | processor, BioMoby, etc. | | License (LGPL) |
| GridAnt | Argonne National Laboratory, USA. http://www-unix.globus.org/cog/projects/gridant/ | Apache Ant, Globus toolkit | GT2, GT3 | Client controllable workflow applications | GTPL |
| GrADS | Collaboration between several American Universities. http://www.hipersoft.rice.edu/grads/ | Globus Tookit, Autopilot, NWS | Globus, Parallel Systems (e.g. MPI) | Compute-intensive and communication-intensive applications with MPI components | Not yet available in public |
| GridFlow | University of Warwick, UK http://www.dcs.warwick.ac.uk/research/hpsg/workflow/workflow.html | Agent-based Resource Management System (ARMS), Performance Analysis and Characterize Environment (PACE) Toolkit, Titan | Parallel Systems (e.g. MPI and PVM) | MPI and PVM based components | Not yet available in public |
| UNICORE | Collaboration between German research institutions and industries http://www.unicore.org/ | UNICORE middleware | UNICORE | Computational-intensive and MPI components | Community Source License |
| Gridbus | The University of Melbourne, Australia. www.gridbus.org | Globus Toolkit | GT2 | Computational- and Data-intensive | GPL |
| Askalon | University of Innsbruck http://dps.uibk.ac.at/askalon | Globus Toolkit | GT2, GT4, WSRF, Web services | Performance-oriented applications | GTPL |

*Table 2*. Workflow Design Taxonomy Mapping.

| Project Name | Structure | Model | Composition Systems | QoS Constraints |
|--------------|-----------|-------|---------------------|-----------------|
| DAGMan | DAG | Abstract | User-directed<br>• Language-based | User specified rank expression for desired resources |

| Project Name | Structure | Model | Composition Systems | QoS Constraints |
|---|---|---|---|---|
| Pegasus | DAG | Abstract | User-directed<br>• Language-based<br>Automatic | N/A |
| Triana | Non-DAG | Abstract | User-directed<br>• Graph-based | N/A |
| ICENI | Non-DAG | Abstract | User-directed<br>• Language-based<br>• Graph-based | Metrics specified by users |
| Taverna | DAG | Abstract/Concrete | User-directed<br>• Language-based<br>• Graph-based | N/A |
| GridAnt | Non-DAG | Concrete | User-directed<br>• Language-based | N/A |
| GrADS | DAG | Abstract | N/A | Estimated application execution time |
| GridFlow | DAG | Abstract | User-directed<br>• Graph-based<br>• Language-based | Application execution time |
| UNICORE | Non-DAG | Concrete | User-directed<br>• Graph-based | N/A |
| Gridbus | DAG | Abstract/Concrete | User-directed<br>• Language-based | N/A |
| Askalon | Non-DAG | Abstract | User-directed<br>• Graph-based<br>• Language-based | Constrains and properties specified by users or pre-defined |

*Table 3.* Workflow Scheduling Taxonomy Mapping.

| Project Name | Planning Scheme | Strategies | Architecture | Decision Making | Performance Estimation |
|---|---|---|---|---|---|
| DAGMan | Just in-time | Performance-driven | Centralized | Local | N/A |
| Pegasus | User-directed, Just in-time | Performance-driven | Centralized | Local, Global | Historical Data Analytical modeling |
| Triana | Just in-time | Based on GAT | Decentralized | Local | N/A |
| ICENI | Prediction-based | Performance & Market-driven | Centralized | Global | Historical Data |

| Project Name | Planning Scheme | Strategies | Architecture | Decision Making | Performance Estimation |
|---|---|---|---|---|---|
| Taverna | Just in-time | N/A | Centralized | N/A | N/A |
| GridAnt | User-directed | N/A | Centralized | N/A | N/A |
| GrADS | Prediction-based | Performance-driven | Centralized | Local, Global | Historical data (empirical) Analytical modeling |
| GridFlow | Simulation-based | Performance-driven | Hierarchical | Local | Analytical modeling |
| UNICORE | User-directed | N/A | Centralized | N/A | N/A |
| Gridbus Workflow | User-directed Just in-time | Market-driven | Hierarchical | Local | N/A |
| Askalon | Just in-time Prediction-based | Performance-driven Market-driven | Decentralized | Global | Analytical modeling Historical data |

*Table 4.* Information Retrieval, Fault-tolerant and Data Movement Taxonomy Mapping.

| Project Name | Information Retrieval | Fault-tolerant | Data Movement |
|---|---|---|---|
| DAGMan | Resource information is retrieved by Condor *Matchmaker* that manages resource and task info advertisement and notification. | Task Level<br>• Migration<br>• Retrying<br>Workflow Level<br>• Rescue workflow | User-directed |
| Pegasus | Resource information retrieved through Globus MDS and RLS. Application component information is retrieved from the GriPhyN Transformation Catalog. | Based on DAGMan | Mediated |
| Triana | Based on GAT protocol | Based on GAT manger | Peer-to-Peer |
| ICENI | Application component information is retrieved by the component metadata service and performance repository service. | Based on middleware | Mediated |
| Taverna | Service information is retrieved through DAML-S web service ontology, domain ontology information service, and UDDI. | Task Level<br>• Retry<br>• Alternate Resource | Centralized |
| GridAnt | Resource information is retrieved through Globus MDS. | N/A | User-directed |

| Project Name | Information Retrieval | Fault-tolerant | Data Movement |
|---|---|---|---|
| GrADS | Resource information is retrieved through Globus MDS and GrADS information service (GIS). Dynamic information is retrieved by NWS. Autopilot is used for provide performance contract information. | Task Level in rescheduling work in GraADS, but not in workflows. | Peer-to-Peer |
| GridFlow | Resource information is retrieved through Titan | Task Level<br>• Alternate resource | Peer-to-Peer |
| UNICORE | UNICORE information service | N/A | Mediated |
| Gridbus | Resource information is retrieved through Grid Market Directory | Task Level<br>• Alternate resource | Centralized |
| Askalon | Static information<br>• Infrastructure-related<br>• Configuration-related<br>• QoS-related<br>Dynamic information<br>• Resource-related<br>• Execution-related | Task Level<br>• Retry<br>• Alternate resource<br>Workflow level<br>• Rescue workflow | Centralized User-directed |

## 3.1 Condor DAGMan

Condor [78][119][115] is a specialized resource management system (RMS) developed at the University of Wisconsin-Madison for compute-intensive jobs. Condor provides a High Throughput Computing (HTC) environment based on large collections of distributed computing resources ranging from desktop workstations to super computers. Condor-G, a component within Condor, utilizes Globus GRAM serving as a uniform interface to heterogeneous batch systems, thus enabling large scale computational Grids. *Matchmaking* within Condor, matches jobs and available resources according to their job and resource classified advertisement. When more than one resource satisfies the job requirement, the resource with higher value of rank expression, which expresses the desirability of a match, is preferred.

The Directed Acyclic Graph Manager (DAGMan) [115][35] is a meta-scheduler for Condor jobs. While Condor aims to discover available machines for the execution of jobs, DAGMan handles the dependencies between the jobs. DAGMan uses DAG as the data structure to represent job dependencies. Each job is a node in the graph and the edges identify their dependencies. Each node can have any number of "parent" or "children" nodes. Children cannot run until their parents have completed. Cycles, where two jobs are both descended from one another, are prohibited, because it would lead to deadlock. DAGMan does not support automatic intermediate data movement, so users have to specify data movement transfer through pre-processing and post-processing commands associated with processing job.

The individual job execution is managed by Condor scheduler. So if a job fails due to the nature of the distributed system, such as loss network connection, it will be recovered by Condor while DAGMan is unaware of such failures. However, DAGMan is responsible for reporting errors for the set of submitted jobs, and generates a rescue DAG. In the case of a job failure, the remainder of the DAG continues until no more progress can be made. A failed node can be retried a configurable number of times. The rescue DAG indicates the uncompleted portions of the DAG with detail of failures. Users can correct the errors of failed jobs and resubmit the rescue DAG.

**3.2 Pegasus in GriPhyN**

GriPhyN [60] aims to support large-scale data management in physics experiments such as high-energy physics, astronomy, and gravitational wave physics. Pegasus [39][40][41] (Planning for Execution in Grids) is a workflow manger in GriPhyN developed by the University of Southern California.

Pegasus performs a mapping from an abstract workflow to the set of available Grid resources, and generates an executable workflow. An abstract workflow can be constructed by querying Chimera [52], a virtual data system, or provided by users in DAX (DAG XML description). An abstract workflow describes the computation in terms of logical files and logical application components and indicates their dependencies in the form of Directed Acyclic Graph (DAG). Before mapping, Pegasus reduces the abstract workflow by reusing a materialized dataset which is produced by other users within a VO. Reduction optimization assumes that it is more costly to produce a dataset than access the processing results. The reduction algorithm removes any antecedents of the redundant jobs that do not have any unmaterialized descendents in order to reduce the complexity of the executable workflow.

Pegasus consults various Grid information services to find the resources, software, and data that are used in the workflow. A Replica Location Service (RLS) [30] and Transformation Catalog (TC) [38] are used to locate the replicas of the required data, and to find the location of the logical application components respectively. Pegasus also queries Globus Monitoring and Discovery Service (MDS) [33] to find available resources and their characteristics.

There are two methods used in Pegasus for resource selection, one is through random allocation, the other is through a performance prediction approach. In the latter approach, Pegasus interacts with Prophesy [67][135], which serves as an infrastructure for performance analysis and modeling of parallel and distributed applications. Prophesy is used to predict the best site to execute an application component by using performance historical data. Prophesy gathers and stores the performance data of every application. The performance information can provide insight into the performance relationship between the application and hardware and between the application, compilers, and run-time systems. An analytical model is produced based on the performance data and is used by the prediction engine to predict the performance of the application on different platforms. It is required that Pegasus send the request associated with information such as the component name, the semantic parameter names and their values, and the list of available resources. The ranking of the given resources is returned by Prophesy after the query is received.

For ease of use, Pegasus is able to generate a workflow from a metadata description of the desired data product with the aid of artificial intelligence planning techniques. Although, the workflow execution of Pegasus is based on static planning and its executable workflow is transformed into Condor jobs for execution management by Condor DAGMan, it has been recently extended to support just in-time scheduling [41] and pluggable task scheduling strategies.

**3.3 Triana**

Triana [117][118] is a visual workflow-oriented data analysis environment developed at Cardiff University. In 2002, Triana was extended to implement a consumer Grid [117] by using a peer-to-peer approach. Recently, Triana has been redesigned and integrated with Grids via GridLab GAT (Grid Application Toolkit) interface [10]. GAT defines a high level API for core Grid service access through JXTA [69], Web services [128], and OGSA [53][121].

Triana provides a visual programming interface with functionality represented by units. Applications are written by dragging the required units onto the workplace and connecting them to construct a workflow. Apart from many implemented tool units, Triana also provides a custom user interface to allow users to build their own units. Several control units (e.g. loop) and logic units (e.g. if) are also provided for users to control the logic of workflow execution. Since control and logic units are implemented as a standard Triana unit, it is easy to introduce new flow patterns. Interconnected units can also be grouped into a group unit, which has the same properties as normal unit.

Triana clients such as Triana GUI can log into a Triana Controlling Service (TCS), remotely build and run a workflow and then visualize the result on their device (e.g. PC, PDA, etc). Each TCS interacts with the Triana engine and every engine provides a service and is capable of executing complete or partial task-graphs locally, or by distributing the code to other servers based on the specified distribution policy for the supplied task-graph. The distribution policy is based on the concept of group units and two distribution policies have been implemented, namely parallel and peer-to-peer. Both policies distribute every unit in the group to separated hosts, however while the peer-to-peer mechanism relies on intermediate data being passed between hosts, there is no such host-based communication with the parallel policy. Since a distributed task-graph is not fixed to a specific set of resources, it can be dynamically allocated to available services in the most effective way.

**3.4 Workflow Management in ICENI**

The ICENI (Imperial College e-Science Network Infrastructure) [88][89] developed at London e-Science Centre provides component-based Grid middleware. Within ICENI, users construct an abstract workflow which is a collection of components and then submit this to ICENI environment for execution.

Each ICENI component is described in terms of meaning, control flow and implementation. The workflow components are primarily composed based on a spatial view, in which all units are represented concurrently, with details of how they relate and interact with each other. Then a temporal view is derived from the spatial view by the system. In the temporal view, workflow information is attached to each component that consists of a graph in which the directed arcs contain the partnership according to the temporal dependence. Within ICENI, the workflow model is similar to that of the YAWL (Yet Another Workflow Language) [4], although simplified in certain respects. The workflow language includes all basic workflow structure such as sequence, parallelism, choice and iteration.

The scheduling service [88] [137][138] within ICENI is responsible for concretizing the abstract workflow. The scheduling task includes matching component meaning with component implementation and mapping these qualified components onto a suitable subset of the available resources. Several scheduling algorithms used to determine resource mapping have been implemented. They include random, best of n random, simulated annealing and game theory. Most schedulers implemented within ICENI aim to provide approximate optimal solutions to map the abstract workflow to a combination of component implementations and resources in terms of execution time and cost. The schedulers take into account all components in applications rather than standalone components. The scheduling framework also allows third-party scheduling algorithms to be plugged in.

ICENI has developed a performance repository system [86] which is able to monitor running applications and obtain and store performance data for the components within the applications. This data is stored within a repository with meta-data about the resource the component was executed on, the implementation of the component used, and the number of other components concurrently running on the same resource. This data can be used by schedulers for future runs of applications to estimate the execution times of each component within the workflow.

Two scheduling schemes [88] are considered within ICENI, namely lazy scheduling and advanced reservation. The metadata of the component implementation indicates which scheme the component can benefit from. Non-reservation component is scheduled to a resource just before it is required, while reservation component has been allocated to a resource and has made a reservation in advance. The schedulers can interrogate the performance repository to predict execution in order to produce accurate reservation. The reservation negotiation protocol is based on WS-Agreement [59].

**3.5 Taverna in [my]Grid**

Taverna [95] is the workflow management system of the [my]Grid [113] project, which aims to exploit Grid technology to develop high-level middleware for supporting personalized in silico experiments in biology. Taverna is a collaboration between several European universities, institutes and industries. The purpose of Taverna is used to assist scientists with the development and execution of bioinformatics workflows on the Grid. Taverna provides data models, enactor task extensions, and graphical user interfaces. FreeFluo [54] is also integrated into Taverna as a workflow enactment engine to transfer intermediate data and invoke services.

In Taverna, data models can be represented in either a graphical format or in an XML based language called Simple Conceptual Unified Flow Language (SCUFL). The data model consists of inputs, outputs, processors, data flow and control flow. In addition to specifying execution order, the control flow can also be trigged by state transitions during the execution of parent processors. Compared to other workflow languages, such as the Business Process Execution Language for Web Services (BPEL4WS) [14] , SCUFL allows implicit iteration over incoming data sets based on a specified strategy. At the execution level, the workflow enactor also provides a multithreading mechanism to speed up the iteration process. Users are allowed to set the *Thread* property to specify how many concurrent instances will send parallel requests to the iteration processor. It is especially suitable for services that are capable of handling significant simultaneous processing, for example, a service that is backed by a cluster. It also can reduce service waiting time since workflow engine can send the next input data at the same time as the service is working on the current input.

Taverna also provides a user-friendly multi-window environment for users to manipulate workflows, validate and select available resources, and then execute and monitor these workflows. The enactment status panel [116] of Taverna shows the current progress of a workflow invocation. It also allows the users to browse the intermediate and final results. Through the enactment panel, users can handle storage of those results on local or remote data stores in a variety of formats.

Fault tolerance [116] in the workflow management of $^{my}$Grid is achieved by setting configuration for each processor in the workflow, for example, the number of retries, time delay and alternate processors. It also allows users to specify the critical level for faults on each processor. If the processor is set as *Critical*, after all retries and alternates have failed, entire workflow execution will be terminated, otherwise, the workflow will continue but children nodes of the failed processor will never be invoked.

$^{my}$Grid follows service-oriented grid architecture and supports several different types of services within the workflow management system, including WSDL-based [133] single operation web services, soaplab bio-services [106] and local services such as programs coded as java classes. In addition, information services such as UDDI (the Universal Description, Discovery and Integration) [122] and ontology directory [134] are adopted for service discovery.

**3.6 GridAnt**

The GridAnt [74][13] is an extensible client-side workflow management system developed by Argonne National Laboratory. It has been designed for Grid end-users as a convenient tool to express and control the execution sequence without having any expertise in sophisticated workflow systems. GridAnt focuses on distributed process management rather than the aggregation of services which is the concern of most other Grid-enabled workflow frameworks.

GridAnt consists of four major components, namely workflow engine, run-time environment, workflow vocabulary and workflow monitoring. The workflow engine is the central controller that handles task dependencies, failure recoveries, performance analysis, and process synchronization. GridAnt workflow engine extends Ant [15], an existing commodity tool for controlling build process in Java, by adding additional components to support workflow orchestration and composition. GridAnt also provides an environment for inter-task communication, so that individual GridAnt tasks can read and write intermediate data by using a globally accessible whiteboard-style communication model. Several important constructs such as constants, arithmetic expressions, global variables, array references, and literals are supported by the run-time environment. GridAnt extends Ant's vocabulary in the Grid domain with the addition of the tags such as grid-copy, grid-authenticate and grid-query. These new tags are used by users to predefine the Grid tasks and construct complex workflows at compile time. It uses a control construct provided by Ant container for expressing parallel and sequential tasks. Furthermore, users are allowed to monitor the progress of the execution by means of graphical visualization tool.

In addition to mapping complex client-side workflows, GridAnt can be used for testing the functionality of different Grid services. It has been developed to support version 2 and version 3 of the Globus toolkit [58] by using the Java CoG kit [73]. It has been applied for Position-Resolved Diffraction [13], which is a new experimental technique for the study of nanoscale structures as part of the Argonne National Laboratory's advanced analytical electron microscope.

**3.7 Workflow management in GrADS**

The Grid Application Development Software (GrADS) project [18] aims to provide programming tools and execution environments for ordinary scientific users to develop, execute, and tune applications on the Grid. GrADS is a collaboration between several American Universities. GrADS supports application development either by assembling domain-specific components from a high-level toolkit or by creating a module by relatively low-level (e.g., MPI ) code [31].

GrADS provides application-level scheduling to map workflow application tasks to a set of resources. New Grid scheduling and rescheduling methods [31] are introduced in GrADS. These scheduling methods are guided by an objective function to minimize the overall job completion time (*makespan*) of the workflow application. The scheduler obtains resource information by using services such as MDS [104] and NWS [131] and locates necessary software on the scheduled node by query GrADS Information Service (GIS). The workflow scheduler ranks each qualified resource for each application component. A rank value is calculated by using "a weighted sum of the expected execution time on the resource and the expected cost of data movement for the component." After ranking, a performance matrix is constructed and used by the scheduling heuristics to obtain a mapping of components onto resources. Three heuristics have been applied in GrADS; those are Min-Min, Max-Min, and Sufferage heuristics [82].

GrADS has built up an architecture-independent model of the workflow component from individual component models. It employs analytical models that are constructed semi-automatically from empirical models (historical data/sample execution data), in order to estimate the performance of a workflow component on a single Grid node. It uses hardware performance counters to collect operation counts from several executions of the workflow components with different, small-size input problems, and then it performs a least-squares fit to the data to construct computational models. In addition, GrADS reuses distance data on small inputs to predict the faction of cache hits and misses on the given data and cache configuration by its memory-hierarchy performance models.

GrADS utilizes Autopilot [102] to monitor performance of the agreement between the application demands and resource capabilities. Once the contract is violated, the rescheduler [31] of the GrADS takes corrective actions. It has been implemented using two rescheduling approaches for MPI applications, the stop/restart approach and process swapping. In the former approach, an executing application component is suspended and migrated to a new resource if better resources are found for improving the execution performance [124]. As a migration event can involve large data transfers, expensive startup costs and significant application code modifications, process swapping provides a lightweight, but less flexible, alternative approach. In process swapping more machines than will actually be used for the computation are launched for an MPI application component, and slower machines in the active set are swapped with faster machines in the inactive set periodically, according to the performance of machines.

**3.8 GridFlow**

GridFlow [25] is a Grid workflow management system developed at the University of Warwick. This work is built on the top of an agent-based resource management system for Grid computing (ARMS) [24]. Rather than focusing on workflow specification and the communication protocol, GridFlow is more concerned about service-level scheduling and workflow management.

There are three layers of Grid resource management within the GridFlow system: the Grid resource, the local Grid and the global Grid. A Grid resource is simply just a particular grid resource; local Grid consists of multiple Grid resources that belong to one organization; and a global Grid consists of all local Grids. Global Grid also provides a portal for compose the workflow.

A workflow in GridFlow is represented as a flow of several different activities, each activity represented by a sub-workflow. Each sub-workflow is a flow of closely related tasks that is to be executed in a local grid. A portal has been developed by GridFlow as graphical user interface for users to compose workflow elements.

The workflow management within GridFlow is conducted by a hierarchical scheduling system including global Grid workflow manager and local Grid sub-workflow scheduling. Global grid workflow manager receives requests from the GridFlow portal with the workflow description in the format of XML, and then simulates workflow execution to find a near-optimal schedule. After the users accept the simulated result, GridFlow schedules the workflow onto different local Grids through ARMS. Within ARMS, each agent represents a local Grid at a global level of Grid resource management, and conducts local Grid sub-workflow scheduling. In contrast to the global Grid workflow management, the local Grid schedulers handle conflicts since scheduled sub-workflows may belong to different workflows.

ARMS has integrated Titan [111], which utilizes performance data obtained from PACE [93], a toolset for resource performance and usage analysis, with iterative heuristic algorithms to minimize the makespan and idle time of a grid resource. PACE can exact control flow, and use an analytical model approach based on queuing theory, to predict application performance on a given set of resources such as time, scalability and system resource usage. Titan also provides Grid resource information.

**3.9 Workflow Management in UNICORE Plus**

UNICORE plus [123] provides seamless and secure access to distributed resources of the German high performance computing centers. UNICORE plus is a follow-on project of UNICORE (Uniform Interface to Computing Resources) [11], started in 1997 to improve uniform interfaces to distributed High Performance Computing and data resources using the mechanisms of the World Wide Web. UNICORE plus provides a programming environment for users to design and execute job flow.

Within UNICORE, one job or job group that can be executed on any UNICORE site may contains other jobs and/or job groups. The original UNICORE job model supports jobs that are constructed as a set of directed acyclic graphs with temporal dependencies. Since UNICORE version 4, advanced flow controls have been added, which include conditional execution (e.g. if-then-else), repeated execution (e.g. do-n), conditional repeated execution (e.g. do-repeat), and conditional suspend action (e.g. hold-job). In addition, three types of run-time conditions are implemented for supporting conditional checking; these are based on the return code of a previous executed task, existence or properties of a file and whether a given time and date have passed.

UNICORE plus provides graphical tools that allow users to create a job flow and convert it into an Abstract Job Object (AJO) which is a serialized java object. The AJO is submitted from a user client to a UNICORE server. The server translates the job specification into a number of batch jobs and dispatches them to the target resource. The server also makes sure that a successor is executed if its predecessors are finished and all necessary data is available at the executing site.

UNICORE allows users to specify jobs and different parts of job group onto multiple resources. The output of individual jobs may be needed by its successors. Therefore, a temporary UNICORE space is created for each job group for transferring data sets. UNICORE also allows users to explicitly specify the transfer function as a task through GUI; it is also able to perform the necessary data movement function without user intervention.

**3.10 Workflow Management in Gridbus**

The Gridbus Toolkit [23] developed by the University of Melbourne provides Grid technologies for service-oriented utility computing. Its architecture is driven by the requirements of Grid economy [22]. A Grid economy mechanism has been proposed as an efficient management technique for distributed resources as it helps in regulating the supply and demand of resources.

The workflow management in Gridbus [139] provides a simple XML-based workflow language for users to define their tasks and dependencies. The workflow description language of Gridbus is aimed towards enabling the expression of parameterization [8] and users' QoS requirement.

The workflow engine of Gridbus provides a hierarchical scheduling architecture to adapt to heterogeneous and dynamic Grid environments. Within the workflow execution engine, the schedules of the workflow tasks are driven by the events by using the tuple space model [55]. An event-driven mechanism with subscription-notification approach makes the workflow execution loosely-coupled and flexible. The system also supports just in-time scheduling, allowing scheduling decision to be made at the time of task execution. The scheduler can also reschedule failed tasks to an alternative resource. In addition, Grid Market Directory (GMD) [140] is utilized by the workflow schedulers for run-time resource discovery.

In contrast to other workflow management systems, the Gridbus workflow system aims towards the utilisation market-based workflow management to Grid environments. The two targeted application domains are natural language processing and molecular modeling.

**3.11 Askalon**
Askalon [48] is a Grid application development and computing environment developed by the University of Innsbruck, Austria. The main objective of Askalon is to simplify the development and optimization of mostly Grid workflow applications that can harness the power of Grid computing.

Askalon comes with two separate composition systems, AGWL [46] and Teuta [47], that support the development of Grid workflow applications. AGWL (Abstract Grid Workflow Language) is an XML-based language. It provides a rich set of constructs to express sequence, parallelism, choice, and iteration workflow structure. In addition, programmers can specify high-level constraints and properties defined over functional and non-functional parameters for tasks and their dependencies which can be useful for a runtime system to optimize the workflow execution. Teuta supports the graphical specification of Grid workflow applications based on the UML activity diagram which is a graphical interface to AGWL.

Askalon provides a new hybrid approach for scheduling workflow applications on the Grid through dynamic monitoring and steering combined with a static optimization. Static scheduling maps entire workflows onto the Grid using genetic algorithms. A problem-independent objective function design allows to plug-in a variety of optimization metrics such as the execution time, efficiency, economical cost, or any user-defined QoS parameter. A dynamic scheduling algorithm takes into consideration the dynamic nature of the Grid resources such as machine crashes or external CPU and network load. Performance contracts are defined for every task and monitor whether tasks execute properly or whether they should be migrated. Askalon develops a fault tolerant execution engine that supports reliable workflow execution in the presence of resource failures through checkpointing and migration techniques.

In order to provide automatic workflow orchestration, Askalon Grid Resource Management (GridARM) provides a distributed GT4-based registry to map generic or domain specific tasks to their implementations. Askalon also includes automatic search for performance problems and faults in Grid infrastructures and applications. The monitoring and performance analysis component provides static information of Grid infrastructure and dynamic information of computational resources, networks, and applications. Dynamic information of workflow-based applications is provided for the entire workflow as well as for invoked applications called within tasks. The performance of workflow components is estimated based on a training phase which measures the actual execution time of tasks for different loads and problem sizes on a variety of Grid sites. The performance estimation of the workflow is conducted based on a combination of historical data obtained from a training phase and analytical modeling.

## 4. Summary and Discussion

We have presented a taxonomy for Grid workflow management systems. The taxonomy focuses on workflow design, workflow scheduling, fault management and data movement. We also survey some workflow management systems for Grid computing and classify them into the different categories using the taxonomy. This paper thus helps to understand key workflow management approaches and identify possible future enhancements.

Many Grid workflow-enabled systems have developed graph-based editing environments. They allow users to compose the workflow by dragging and dropping components on a composition panel. A workflow abstract specification or concrete specification is then generated by these visual tools and passed to the workflow enactment engine. These processes are transparent to users for better usability. Currently, only Pegasus supports automatic workflow composition. In order to support the automatic composition, catalogs with rich information about application components and services need to be addressed. Besides GriPhyN Chimera system and UDDI (Universal Description, Discovery and Integration) directory service for web services discovery, many efforts from semantic Web such as DAML+OIL ontology [66] can be used for providing accurate description and flexible discovery of application components and services.

Most of the Grid workflow projects discussed in this paper have their own graphical workflow modeling and language. Obviously, the lack of standardized syntax and semantic description for workflow modeling and language results in many replicated works. More effort is thus needed towards workflow modeling standardization. Even though there are some proposed workflow languages for web services such as BPEL4WS, they are still not sufficient due to lack of implementation, levels of abstraction and limited supported services [9].

Quality of Service (QoS) issues have not been addressed very well in most Grid workflow management systems due to the use of system centric policies in resource allocation. However, once the commercial-oriented workflow management systems come into view, supporting QoS will become increasingly critical at both specification level and execution level. At the specification level, workflow languages need to allow users to express their QoS requirements. At the execution level, the workflow scheduling must be able to map the workflow onto Grid resources to meet users' QoS requirements. Therefore, the role of market-driven strategies will become increasingly important, currently being ignored in most Grid workflow management systems. Trust-based scheduling is another approach to improve QoS in open distributed systems such as Grid and peer-to-peer; however, it has not been addressed very well in the context of workflow management.

It is impossible to make an optimal scheduler without knowledge of estimated time of task execution. Several performance information services are utilized in Grid workflow projects to predict performance prediction. One example is PACE employed in GridFlow project. It uses analytical model to predict application performance, but the current implementation is only adapted to MPI program. Prophesy used by Pegasus uses historical performance database to gain insight into the relationship between applications and resources in order to predict the performance of the applications on a given set of resources. Similarly, ICENI developed a performance repository system which is able to collect performance data for application components. GrADS have developed two analytical models for their GrADS programs.

Given the dynamic nature of Grid environments, fault tolerance should be fully supported by Grid workflow management systems. However, most fault handling techniques have not been developed or implemented in many Grid workflow systems, especially at the workflow execution level. It is hard for a workflow management system to survive in real Grid environments without robust fault handling techniques.

**ACKNOWLEDGEMENTS**

We would like to acknowledge all developers of the workflow management systems described in the paper. We also want to thank Chee Shin Yeo, Hussein Gibbins, Anthony Sulistio, Srikumar Venugopal, Baden Hughes (Melbourne University), Rob Gray (Monash University, Australia), Wolfram Schiffmann (FernUniversitaet in Hagen, Germany), Ivona Brandic (University of Vienna, Austria), Soonwook Hwang (National Institute of Informatics, Japan), Ewa Deelman (University of Southern California), and Chris Mattmann (NASA Jet Propulsion Laboratory, USA), Henan Zhao (University of Manchester, UK), Thomas Fahringer (University of Innsbruck, Austria), Ken Kennedy, Anirban Mandal, and Chuck Koelbel (Rice University, USA) for their comments on this paper. This work is partially supported through the Australian Research Council (ARC) Discovery Project grant and Storage Technology Corporation sponsorship of Grid Fellowship.